# Unexpected persistence of cis-bridged chains in compressed AuF$_3$

Dominik Kurzydłowski,*[a,b] Serhiy Kobyakov,[a] Zoran Mazej,[c] Sharad Babu Pillai, [d] Brahmananda Chakraborty, [e] Prafulla K. Jha [d]

[a.] *Faculty of Mathematics and Natural Sciences, Cardinal Stefan Wyszyński University, Warsaw 01-038 , Poland*

[b.] *Centre of New Technologies, University of Warsaw, ul. Banacha 2c, Warsaw 02-097, Poland*

[c.] *Department of Inorganic Chemistry and Technology, Jožef Stefan Institute, SI-1000 Ljubljana, Slovenia*

[d.] *Department of Physics, Faculty of Science, The Maharaja Sayajirao University of Baroda, Vadodara 390002, India*

[e.] *High Pressure and Synchrotron Radiation Physics Division, Bhabha Atomic Research Centre, Trombay, Mumbai 400085, India*

**Raman scattering measurements indicate that cis-bridged chains are retained in AuF$_3$ even at a compression of 45 GPa - in contrast to meta-GGA calculations suggesting that structures with such motifs are thermodynamically unstable above 4 GPa. This metastability implies that novel gold fluorides (e.g. AuF$_2$) might be attainable at lower pressures than previously proposed.**

Due to its chemical inertness gold forms connections mainly with highly reactive elements. In fluorine-rich compounds gold usually adopts the +3 oxidation state,[1] although compounds with gold(I),[2] gold(II),[3,4] and gold(V)[5,6] are also known. Only two binary gold fluorides are stable in the solid state: AuF$_3$ [7] and AuF$_5$.[5] Both AuF and AuF$_2$ can only be stabilized as isolated molecules in a collision-free environment.[8,9] Organometallic compounds containing gold both in the +1 and +3 oxidation state are important catalysts in a range of reactions,[10] and AuF$_3$ is a promising precursor for the synthesis of extremely Lewis acidic catalysts.[11,12]

Theoretical studies hint at the increased reactivity of fluorine at high pressure (p > 1 GPa).[13–20] This creates an opportunity for the stabilization of novel gold fluorides, as indicated by some early studies.[21–23] Indeed, recent calculations suggest that at large compression AuF$_2$, AuF$_4$, and even AuF$_6$ become thermodynamically stable.[19,20] However, these predictions still await experimental verification. In no fact high- pressure experimental data has been reported even for the most stable gold fluoride, AuF$_3$.

In an attempt to understand how large compression affects the properties of AuF$_3$, we performed a study combining both experimental (Raman spectroscopy) and theoretical (Density Functional Theory, DFT) methods. At ambient condition AuF$_3$ adopts a hexagonal structure (space group *P6$_1$22, Z* = 6)[24] comprised of cis-bridged helical AuF$_3$ chains built from planar AuF$_4$ squares (Fig. 1a). Apart from two pairs of Au-F bonds (bridging: $R_b$ = 2.00 Å, terminal: $R_t$ = 1.88 Å) there are two equivalent inter-chain Au···F contacts ($R_{inter}$ = 2.76 Å). The square planar coordination of $Au^{3+}$ ($d^8$ electron count) leads to diamagnetic properties of $AuF_3$; the orange colour of this compound indicates presence of an electronic band gap between 2.5 and 2.9 eV.

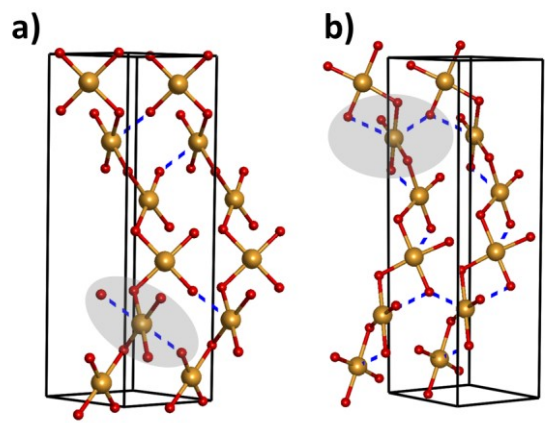

**Fig. 1** The ambient-pressure $P6_122$ structure of $AuF_3$ (a) together with the high-pressure $P6_1$ polymorph (b). Gold/red balls mark Au/F atoms. Grey lobes mark differences in secondary Au···F contacts (depicted with dashed blue lines)

The non-centrosymmetric $P6_122$ structure exhibits 28 Raman-active modes ($5A_1 + 12E_2 + 11E_1$). By comparing the ambient-pressure Raman spectrum with frequencies and intensities of Raman bands simulated with DFT (Supplementary Information, SI, Table S1 and Fig. S1) we were able to assign all of the 15 bands observed in experiment ($3A_1 + 5E_2 + 7E_1$). The high frequency modes can be identified as Au-F stretches: terminal (Au-$F_t$) above 600 $cm^{-1}$ and bridging (Au-$F_b$) between 430 and 550 $cm^{-1}$.

Upon compression several changes in the Raman spectrum can be observed starting from 1.5 GPa (Fig. 2). These include softening of the stretching modes, with the exception of the two lowest ones ($E_2$ and A). Moreover new bands appear and gain in intensity upon compression, in particular one Au-$F_t$ and one Au-$F_b$ stretching mode (SI, Fig. S2). These changes mark a transition from $P6_122$ to a related structure of $P6_1$ symmetry (*vide infra*). This notion is supported by an excellent agreement between the band positions simulated for this structure and those observed experimentally (Fig. 2). This agreement allows us to assign all of the 21 bands observed experimentally (the $P6_1$ structure exhibits a total of 34 Raman-active modes: $11A + 12E_2 + 11E_1$). Good agreement between experiment and theory is also found when comparing the intensities of Raman bands (SI, Fig. S3). Our measurements do not reveal any other phase transitions up to 45 GPa.

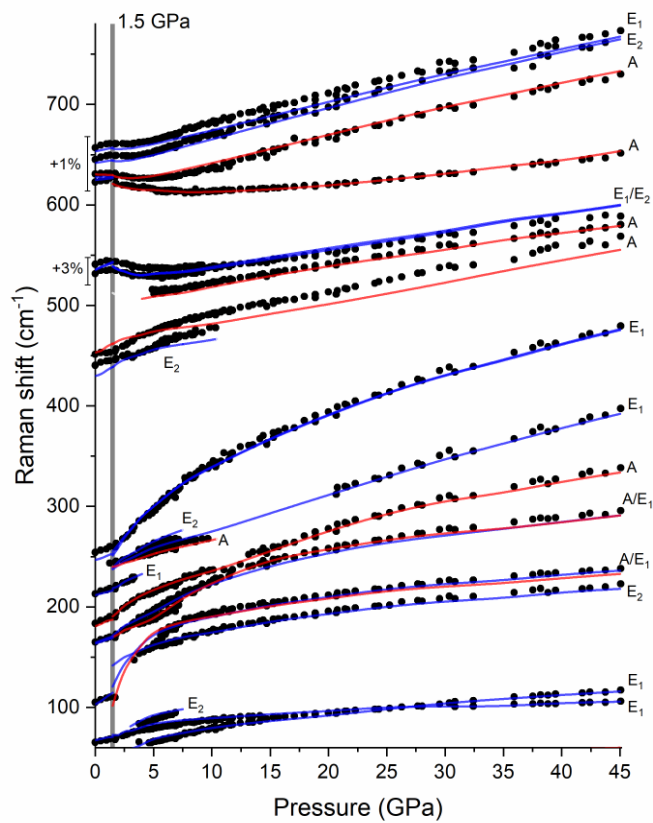

**Fig. 2** Pressure dependence of the frequencies of the Raman bands of solid $AuF_3$ measured upon compression (black points). Lines mark frequencies simulated for the $P6_122$ structure (below 1.5 GPa) and for the $P6_1$ structure (above 1.5 GPa) with the SCAN functional (blue/red lines mark modes of A/E symmetry, symmetry labels are given for the $P6_1$ structure). Theoretical frequencies were scaled only for the six highest frequency modes (by +1 % and +3 % as indicated).

Calculations indicate that the $P6_122 \rightarrow P6_1$ transition should occur at 1 GPa (Fig. 3a), further supporting our assignment of the experimental data. This transition is of second order and is driven by the softening of an $A_2$ mode of the $P6_122$ structure (calculations indicate that this mode should become imaginary at 1 GPa). The $P6_1$ polymorph retains the cis-bridge chains present in $P6_122$, but with a different secondary Au(III) coordination – due to bending within the chains one of the Au⋯F inter-chain contacts is replaced by an intra-chain contact (Fig 1b). Interestingly, $P6_1$ exhibits negative linear compressibility of the *c* cell vector up to approximately 23 GPa (Fig. 3b). Calculations indicate also that one of the terminal and one of the bridging Au-F bonds lengthen upon initial compression (Fig. 3c). This behaviour explains the softening of the Au-F stretching modes observed in the Raman scattering experiment.

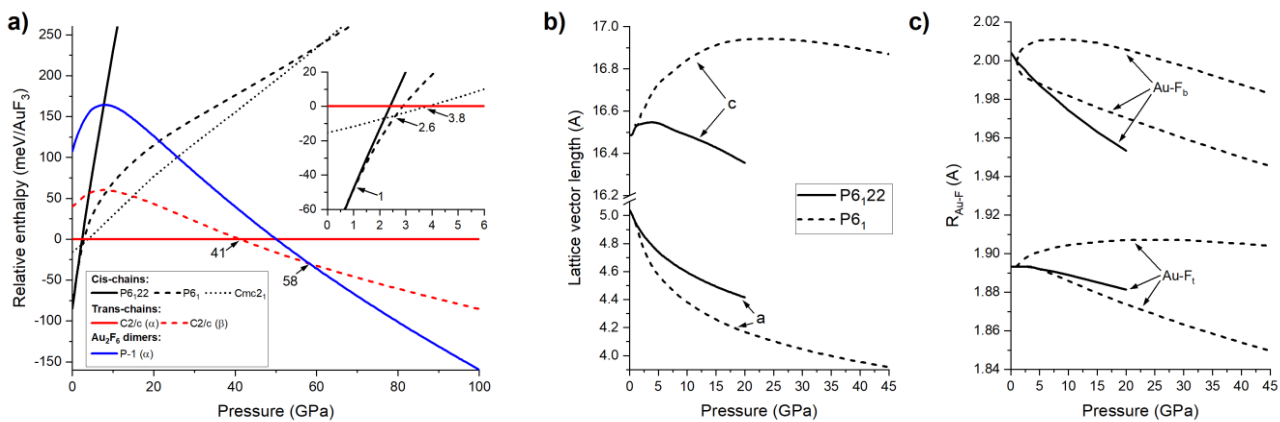

**Fig. 3** Pressure dependence of the relative enthalpy (referenced to that of *C2/c (α)*) of the high-pressure polymorphs of AuF$_3$ (a); pressures at which phase transitions should occur are indicated in GPa. The pressure dependence of the lattice vectors (b) and Au-F bond lengths (c) calculated for the *P6$_1$22* (solid lines) and *P6$_1$* (dashed lines) structures.

It is noteworthy to point that the *P6$_1$* structure was not reported in previous theoretical studies. In the work of Lin et al. the *P6$_1$22* polymorph was predicted to transform into a *P–1* structure containing Au$_2$F$_6$ dimers (denoted *P–1(α)*).[19] A more recent study proposed that *P6$_1$22* should transform into a BrF$_3$-type structure[25] (*Cmc2$_1$* symmetry) at 6 GPa with a subsequent transition into *P–1(α)* at 25 GPa.[20]

In order to elucidate the structure preferences of compressed AuF$_3$ we performed extensive evolutionary algorithm searches for the lowest-enthalpy structures of this compound. We identified two novel structures of *C2/c* symmetry (*C2/c(α)* with Z = 4 and *C2/c(β)* with Z = 8). In contrast to the previously studied structures these polymorphs contain trans-linked AuF$_3$ chains (SI, Fig. S4).

Comparison of the enthalpy of various AuF$_3$ phases (Fig. 3a) reveals a rich phase diagram up to 100 GPa:

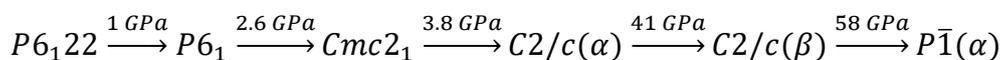

$$P6_122 \xrightarrow{1\ GPa} P6_1 \xrightarrow{2.6\ GPa} Cmc2_1 \xrightarrow{3.8\ GPa} C2/c(\alpha) \xrightarrow{41\ GPa} C2/c(\beta) \xrightarrow{58\ GPa} P\bar{1}(\alpha)$$

The above phase transition sequence indicates that structures built from cis-chains (*P6$_1$22*, *P6$_1$*, *Cmc2$_1$*) are only stable at low pressure (p < 4 GPa). At moderate pressures (4 to 60 GPa) polymorphs exhibiting trans-chains (*C2/c(α)*, *C2/c(β)*) are thermodynamically most stable. Finally above 60 GPa a polymorph containing Au$_2$F$_6$ dimers (analogous to Au$_2$Cl$_6$ units found in AuCl$_3$)[26] becomes the ground state structure of AuF$_3$. Interestingly, despite large changes in the bond connectivity the transition from trans-chain *C2/c(β)* to dimeric *P–1(α)* is associated with a volume decrease below 1 % (Fig. 4a). We also note that in all of the high-pressure phases gold retains its square coordination with Au-F bonds not exceeding 2 Å (additional Au···F contacts are at least 15 % longer). As a result AuF$_3$ should remain diamagnetic even at 100 GPa. Despite a pressure-induced decrease of the electronic band gap observed for all of the high-pressure phases (Fig. 4b), this compound should also retain its insulating properties at this pressure.

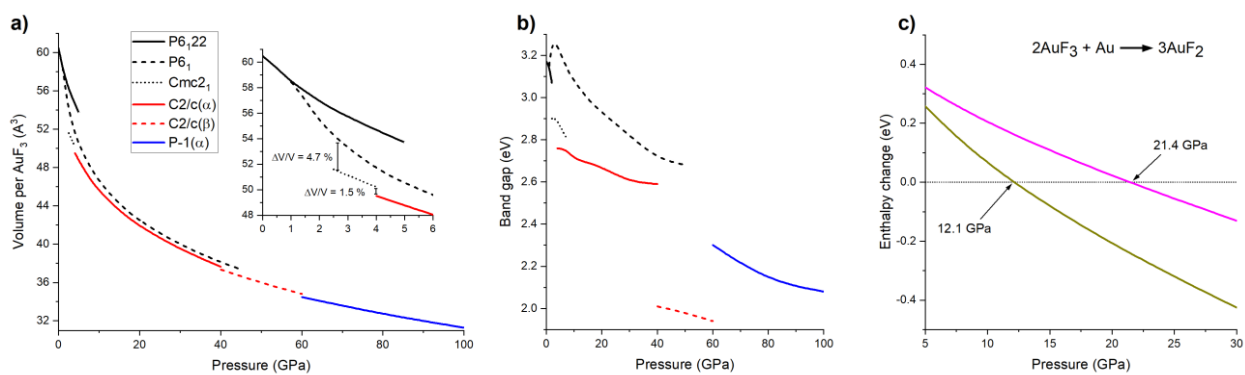

**Fig. 4** Pressure dependence of the calculated volume (a) and the electronic band gap (b) of $AuF_3$. The enthalpy change of the reaction: $2AuF_3 + Au \rightarrow 3AuF_2$ is given in (c). This change was calculated assuming that $AuF_2$ adopts the *Pnma* structure,[20] while $AuF_3$ adopts either the thermodynamically most stable structure (*C2/c(α)*, magenta line) or the experimentally observed *P6₁* structure (dark yellow line). The pressure at which $AuF_2$ becomes thermodynamically stable is indicated for both scenarios.

Calculations hint that the *P6₁* phase has only a narrow window of thermodynamic stability (1 – 2.6 GPa). The fact that experimentally it is observed up to 45 GPa must be attributed to its metastability. This resembles the situation found in other fluorides(e.g. $ZnF_2$, $MgF_2$)[27,28] where metastable structures are formed at large compression. While the *P6₁22* → *P6₁* transition is driven by a phonon instability and hence does not require any activation energy, a considerable energetic barrier is expected for the transition of *P6₁* into *C2/c(α)* due to large energetic difference between cis and trans chain arrangements.[29] The *P6₁* → *C2/c(α)* phase transition might be observed at high temperatures and pressures, however such conditions might induce reactions between $AuF_3$ and the diamond/gasket of the DAC.

The observed metastability of $AuF_3$ has important implication for the possibility of obtaining novel fluorides of gold at large compression. As shown in Fig. 4c the persistence of $AuF_3$ in the high-enthalpy *P6₁* structure results in a lowering of the pressure required for the stabilization of $AuF_2$ by over 9 GPa. The findings of this study, in particular the possibility of obtaining metastable polymorphs at high pressure can prove important in other field of the high-pressure chemistry of gold, for example the Au-O binary phases.[30–32]

The authors acknowledge the support from the Polish National Science Centre (NCN) within grant no. UMO-2014/13/D/ST5/02764. This research was carried out with the support of the Interdisciplinary Centre for Mathematical and Computational Modelling (ICM), University of Warsaw, under grants no. GB74-8 and GA67-13. We thank Marek Tkacz for sharing the laser drilling system at the Institute of Physical Chemistry of the Polish Academy of Sciences. SBP acknowledges financial support from Council of Scientific and Industrial Research (CSIR), Govt of India in the form of Senior Research Fellowships and Ministry of Earth Sciences (MoES), Govt of India. B. Chakraborty would like to thank the staff of BARC computer division for access to its super-computing facility.

**Experimental and computational details**

**Raman spectroscopy:** Raman spectra were acquired at room temperature with the Alpha300M+ confocal microscope (Witec Gmbh) equipped with a motorized stage. We used a 532 nm laser line delivered to the microscope through a single-mode optical fiber. In order to avoid laser-induced decomposition of $AuF_3$ the lase power at the sample did not exceed 2 mW. The Raman signal was collected through a 20× long working distance objective, and passed through a multi-mode optical fiber (50 μm core diameter) to a lens-based spectrometer (Witec UHTS 300, f/4 aperture, focal length 300 mm) coupled with a back-illuminated Andor iDUS 401 detector thermoelectrically cooled to -60°C. The spectra were collected with the use of a 1800 mm grating resulting in a 1.2 cm$^{-1}$ spectral resolution. Typical acquisition times ranged from 1 to 5 s with 20 to 30 accumulations. The spectra were post-processed (background subtraction and cosmic-ray removal) with the Project FIVE software (Witec Gmbh). The position of Raman bands was established with the Fityk 1.3.1 software by fitting the observed bands with Pseudo-Voigt profiles.[33]

**High-pressure experiments:** A total of six high-pressure runs were conducted with the use of a diamond anvil cell (DAC) equipped with low-fluorescence Ia diamonds with a 500 μm culet (bevelled from 600 μm) and a stainless-steel gasket pre-indented to a thickness of 35 μm. The gasket hole with a radius of 120 μm was laser-drilled. The hole was filled in an Ar-atmosphere glovebox by powdered gold trifluoride synthesized as described below. The pressure inside the cell was determined from the shift of the R1 ruby fluorescence line.[34] In those cases when the signal from ruby could not be acquired the pressure was determined from the shift of the first-order Raman peak of the diamond anvil tip.[35] $AuF_3$ decomposition or possible reaction with the diamonds, ruby, and gasket were ruled out by performing Raman mapping (2D scans) of the entire sample at selected pressures.

**Synthesis of $AuF_3$:** Gold(III) fluoride was prepared by fluorination of $AuCl_3$ with elemental fluorine in anhydrous hydrogen fluoride. In a typical reaction, 2 g of $AuCl_3$ was weighed into reaction vessel ($V$ = 35 ml) inside a dry-box. Anhydrous HF (7 ml) was condensed onto the reaction mixture at 77 K and the reaction vessel was warmed to ambient temperature. Elemental fluorine was slowly added to the reaction vessel until the pressure of 4 bar was attained. The reaction mixture was allowed to stir at ambient temperature. After one day, the volatiles were partially pumped away until the pressure in the reaction vessel was equal to that of vapour pressure of liquid HF at ambient temperature. After the reaction mixture was warmed to ambient temperature again, the new portion of fluorine was added again . The whole procedure has been repeated two more times. With the last portion of fluorine, which was already in an excess, the reaction mixture was left for four days.

**DFT calculations:** Periodic DFT calculations of the geometry and enthalpy of various polymorphs of $AuF_3$ up to 100 GPa utilized the SCAN meta-GGA functional.[36] We found it to reproduce very well the geometry and vibrational frequencies of the ambient-pressure structure ($P6_122$ symmetry)[24] of $AuF_3$ (see Table S1). The projector-augmented-wave (PAW) method was used in the calculations,[37] as implemented in the VASP 5.4 code.[38,39] The cut-off energy of the plane waves was set to 800 eV with a self-consistent-field convergence criterion of $10^{-6}$ eV. Valence electrons (Au: $5d^{10}$, $6s^1$; F: $2s^2$, $2p^5$) were treated explicitly, while standard VASP pseudopotentials, accounting for scalar relativistic effects were used for the description of core electrons. The $k$-point mesh spacing was set to $2\pi \times 0.04$ Å$^{-1}$. All structures were optimized until the forces acting on the atoms were smaller than 5 meV/Å.

Evolutionary algorithm searches were performed in order to identify candidates for high-pressure phases of $AuF_3$. For this we used the XtalOpt software (version r12)[40] coupled with periodic DFT calculations utilizing the PBE functional.[41] These searches were conducted at 10/40/80 GPa for $Z$ = 2, 4, and 6. Thermodynamic stability of various $AuF_3$ polymorphs was judged by comparing their enthalpy ($H$), and thus the calculations formally correspond to $T$ = 0K at which the Gibbs free energy ($G = H - S \cdot T$, where $S$ is the entropy) is equal to $H$.

Calculations of Γ-point vibration frequencies were conducted in VASP 5.4 utilizing the SCAN functional. The finite-displacement method was used with a 0.007 Å displacement, and a tighter SCF convergence criterion ($10^{-8}$ eV). In case of the $P6_1$ and $P6_122$ structures we additionally calculated the intensity of Raman-active modes using density-functional perturbation theory (DFPT),[42] as implemented in the CASTEP code (academic release version 19.11).[43] In these calculations the LDA approximation was used together with norm-conserving pseudopotentials (cut-off energy 940 eV). The calculations yielded the Raman activity of each vibrational mode ($S_i$) from which the relative intensity could be estimated assuming that the intensity of the Raman band is proportional to the following factor[44]:

$$\frac{(v_0 - v_i)^4}{v_i \left(1 - e^{-hv_i c / kT}\right)} S_i$$

where $v_0$ is the laser frequency, $v_i$ is the mode frequency, T is the temperature.

For calculations of the electronic band gap ($E_g$) of the most stable structure of $AuF_3$ we employed the Heyd-Scuseria-Ernzerhof (HSE06) functional,[45] which is a hybrid functional mixing the GGA functional of Perdew et al.,[41] with 25% of the Hartree-Fock exchange energy. Calculations were performed for the SCAN-relaxed structures.

Visualization of all structures was performed with the VESTA software package.[46] For symmetry recognition we used the FINDSYM program.[47] Input geometries for CATEP calculations were generated with CIF2Cell.[48] Group theory analysis of the vibrational modes was performed with the use of the Bilbao Crystallographic Server.[49]

# SUPPLEMENTARY INFORMATION

## Tables

**Table S 1** Comparison of the experimental geometry and frequencies of the Raman-active vibrational modes of the ambient pressure structure of AuF$_3$ (space group $P6_122$) with data obtained from calculations utilizing the SCAN functional. Cell vectors and Au-F distances are in Å, volume in Å$^3$, frequencies in cm$^{-1}$.

| | Crystal structure | | Raman-active vibrations | | |
|---|---|---|---|---|---|
| | Exp. (ref. [24]) | SCAN (this work) | SCAN (this work) | Exp. (this work) | Exp (ref. [50]) |
| **a** | 5.1508 | 5.0410 (–2.1 %) | E$_2$ | 32 | |
| **c** | 16.2637 | 16.4871 (+1.4 %) | A$_1$ | 34 | |
| **V** | 373.68 | 362.84 (–2.9 %) | E$_1$ | 43 | |
| **Au-F$_t$** | 1.876 | 1.893 (+0.9 %) | E$_2$ | 59 | |
| **Au-F$_b$** | 2.000 | 2.004 (+0.2 %) | E$_1$ | 67 | 65 | 66 |
| | | | E$_2$ | 93 | |
| | | | E$_1$ | 103 | 105 | |
| | | | E$_2$ | 131 | |
| | | | E$_1$ | 164 | 165 | 164 |
| | | | E$_2$ | 164 | |
| | | | A$_1$ | 181 | 183 | 182 |
| | | | E$_1$ | 183 | |
| | | | E$_2$ | 212 | 213 | |
| | | | E$_1$ | 231 | |
| | | | E$_2$ | 233 | |
| | | | A$_1$ | 237 | |
| | | | E$_2$ | 241 | |
| | | | E$_1$ | 247 | 254 | 254 |
| | | | E$_2$ | 430 | 440 | 436 |
| | | | E$_1$ | 446 | |
| | | | A$_1$ | 453 | 451 | |
| | | | E$_1$ | 520 | 531 | |
| | | | E$_2$ | 520 | 541 | 540 |
| | | | E$_1$ | 619 | 623 | 622 |
| | | | E$_2$ | 621 | | |
| | | | A$_1$ | 624 | 631 | 631 |
| | | | E$_2$ | 636 | 645 | 644 |
| | | | E$_1$ | 647 | 657 | 655 |

# Figures

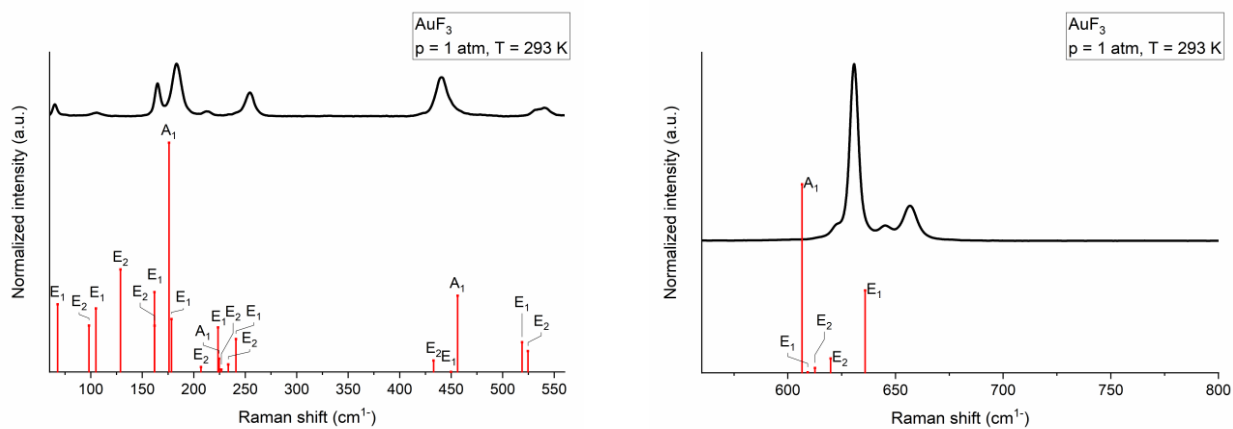

**Fig. S 1** Comparison of the experimental Raman spectrum of $AuF_3$ at 1 atm (black lines) with Raman intensities simulated for the $P6_122$ structure with LDA (red bars). Labels denote symmetry of each mode.

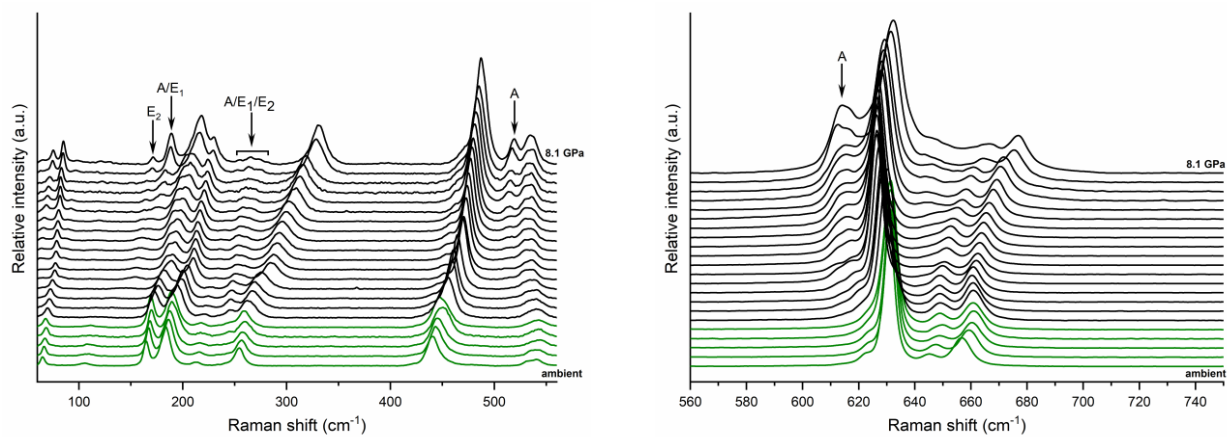

**Fig. S 2** Evolution of the Raman spectrum of powdered $AuF_3$ from ambient pressure to 8.1 GPa (spectra are offset for clarity). Green lines denote spectra corresponding to the $P6_122$ structure, black to the $P6_1$ polymorph. The bands originating from the latter structure which are gaining in intensity upon compression are marked with their symmetry.

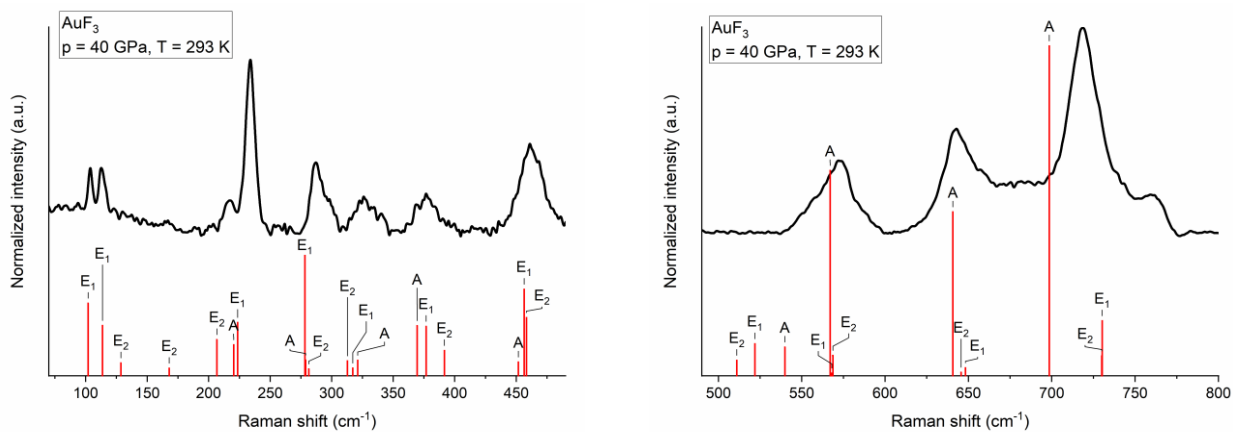

**Fig. S 3** Comparison of the experimental Raman spectrum of $AuF_3$ at 40 GPa (black lines) with Raman intensities simulated for the $P6_1$ structure with LDA (red bars). Labels denote the symmetry of each mode.

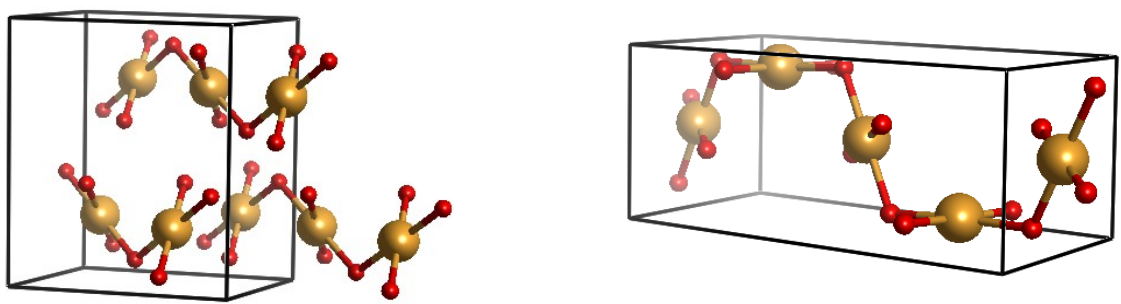

**Fig. S 4** The *C2/c(α)* (left) and *C2/c(β)* (right) structures of AuF$_3$.

# Crystal structure of the novel high-pressure polymorphs of AuF$_3$

Structures optimized with SCAN and given below in VASP format.

**P61 10 GPa**
```
  1.00000000000000
    4.3748027567666048    0.0000019211749695    0.0000000000000000
   -2.1874030421696307    3.7886893634311019    0.0000000000000000
    0.0000000000000000    0.0000000000000000   16.8583856127128193
  Au   F
   6   18
Direct
  0.3738400234095351  0.3746267678221855  0.2563806795449395
  0.6253732041778122  0.9992132735873582  0.5897140605449422
  0.0007867874126433  0.6261599885904587  0.9230473555449450
  0.6261599525904629  0.6253731851778142  0.7563807085449383
  0.3746267598221849  0.0007867624126447  0.0897140605449422
  0.9992132195873609  0.3738400264095390  0.4230473845449438
  0.9183090076980349  0.7538152070911295  0.0324423391489788
  0.2461847939088671  0.1644937776069071  0.3657756801489782
  0.8355062733930936  0.0816910303019611  0.6991090041489798
  0.0816910333019649  0.2461847919088669  0.5324423571489802
  0.7538151990911288  0.8355062373930977  0.8657755961489784
  0.1644937616069058  0.9183089846980366  0.1991090041489798
  0.5795384982248564  0.0754251411965683  0.9880205002302134
  0.9245749418034279  0.5041133720282787  0.3213538802302125
  0.4958866259717212  0.4204615097751443  0.6546872042302141
  0.4204615017751436  0.9245749138034256  0.4880205562302109
  0.0754251471965688  0.4958866429717190  0.8213537952302090
  0.5041133970282843  0.5795385362248524  0.1546871902302129
  0.5873155901351055  0.8312520331142821  0.3034564560758710
  0.1687480168857221  0.7560635850208328  0.6367897800758655
  0.2439364379791726  0.4126844438648973  0.9701231880758669
  0.4126844248648922  0.1687479978857240  0.8034564270758651
  0.8312520181142773  0.2439364299791649  0.1367897800758655
  0.7560635720208353  0.5873156031351030  0.4701231040758671
```

**C2_c_a (pr) 10 GPa**
```
  1.00000000000000
    3.9035752866891102   -0.1189848993802109   -0.1431777998491730
   -0.6463259945999539    4.9105455350708587    0.0167703409590615
   -0.6518440924924354   -1.0240517832483966    4.8018326821234263
```

Au   F
   2    6
Direct
 0.0590415299367161  0.8574544021449929  0.7656881333340257
 0.5589981354357004  0.3574183597412787  0.2657274150614742
 0.1477513107966659  0.1343211949633632  0.3313714930518468
 0.6477844523901057  0.9231207036413080  0.5425929337264603
 0.4702632668318504  0.7917818999196079  0.9887875195795321
 0.8089754098535858  0.4623667724644897  0.6608608831946006
 0.9702739609864771  0.5805041857644925  0.2001097438223262
 0.3090223137688882  0.2525720013604511  0.8706040982297689

**C 2_c_b 30 GPa**
  1.00000000000000
     6.4691300314068805    0.0000000000000000    0.0234967178343975
     0.0000000000000000    4.2616545207618222    0.0000000000000000
    -3.3975034015437315    0.0000000000000000   11.3354807883680113
  Au   F
   8   24
Direct
 0.9826028587195623  0.4999012084448681  0.4701376631334263
 0.9826028587195623  0.5000987915551319  0.9701376331334236
 0.4826028587195624  0.9999012084448681  0.4701376631334263
 0.4826028587195624  0.0000987915551319  0.9701376331334236
 0.9825117346564328  0.1265272024165468  0.7201066694421031
 0.9825117346564328  0.8734728115834544  0.2201066694421031
 0.4825117046564304  0.6265271884165456  0.7201066694421031
 0.4825117046564304  0.3734728115834543  0.2201066694421031
 0.1558605302952651  0.3778928610278049  0.3720176406896429
 0.1558605302952651  0.6221071089721998  0.8720176406896428
 0.6558605302952650  0.8778928910278002  0.3720176406896429
 0.6558605302952650  0.1221071389721951  0.8720176406896428
 0.0206514844785179  0.1107108543295730  0.8902659252085585
 0.0206514844785179  0.8892891456704269  0.3902659252085585
 0.5206514604785158  0.6107108543295731  0.8902659252085585
 0.5206514604785158  0.3892891456704270  0.3902659252085585
 0.9444800291409980  0.1104873422375776  0.5499710422769999
 0.9444800291409980  0.8895126727624272  0.0499710422769999
 0.4444800291409980  0.6104873272375728  0.5499710422769999
 0.4444800291409980  0.3895126727624271  0.0499710422769999
 0.7879052744679648  0.3692862034649759  0.2669621208724425
 0.7879052744679648  0.6307138255350229  0.7669621208724425
 0.2879052744679648  0.8692861744649771  0.2669621208724425
 0.2879052744679648  0.1307137965350240  0.7669621208724425
 0.3093233611102870  0.1218855177450941  0.5682679744665410
 0.3093233611102870  0.8781144892549030  0.0682679744665409
 0.8093233611102869  0.6218855107450970  0.5682679744665410
 0.8093233611102869  0.3781144892549030  0.0682679744665409
 0.6771447021309762  0.1307000862975800  0.6733109659102802
 0.6771447021309762  0.8692998907024215  0.1733109659102803
 0.1771447021309762  0.6307001092975785  0.6733109659102802
 0.1771447021309762  0.3692999207024241  0.1733109659102803

# References

1   F. Mohr, *Gold Bull.*, 2004, **37**, 164–169.

2   I. C. Hwang, S. Seidel and K. Seppelt, *Angew. Chemie Int. Ed.*, 2003, **42**, 4392–4395.

3   S. H. Elder, G. M. Lucier, F. J. Hollander and N. Bartlett, *J. Am. Chem. Soc.*, 1997, **119**, 1020–1026.

4   S. Seidel and K. Seppelt, *Science*, 2000, **290**, 117–118.

5   I.-C. Hwang and K. Seppelt, *Angew. Chemie Int. Ed.*, 2001, **40**, 3690–3692.

6   J. H. Holloway and G. J. Schrobilgen, *J. Chem. Soc., Chem. Commun.*, 1975, 623–624.

7   F. W. B. Einstein, P. R. Rao, J. Trotter and N. Bartlett, *J. Chem. Soc.*, 1967, 478.

8   C. J. Evans and M. C. L. Gerry, *J. Am. Chem. Soc.*, 2000, **122**, 1560–1561.

9   X. Wang, L. Andrews, K. Willmann, F. Brosi and S. Riedel, *Angew. Chemie Int. Ed.*, 2012, **51**, 10628–10632.

10  J. Miró and C. del Pozo, *Chem. Rev.*, 2016, **116**, 11924–11966.

11  M. A. Ellwanger, S. Steinhauer, P. Golz, T. Braun and S. Riedel, *Angew. Chemie Int. Ed.*, 2018, **57**, 7210–7214.

12  M. A. Ellwanger, C. von Randow, S. Steinhauer, Y. Zhou, A. Wiesner, H. Beckers, T. Braun and S. Riedel, *Chem. Commun.*, 2018, **54**, 9301–9304.

13  M. Miao, *Nat. Chem.*, 2013, **5**, 846–852.

14  J. Botana, X. Wang, C. Hou, D. Yan, H. Lin, Y. Ma and M. Miao, *Angew. Chemie Int. Ed.*, 2015, **54**, 9280–9283.

15  D. Kurzydłowski and P. Zaleski-Ejgierd, *Sci. Rep.*, 2016, **6**, 36049.

16  D. Luo, Y. Wang, G. Yang and Y. Ma, *J. Phys. Chem. C*, 2018, **122**, 12448–12453.

17  D. Luo, J. Lv, F. Peng, Y. Wang, G. Yang, M. Rahm and Y. Ma, *Chem. Sci.*, 2019, **10**, 2543–2550.

18  J. Lin, Z. Zhao, C. Liu, J. Zhang, X. Du, G. Yang and Y. Ma, *J. Am. Chem. Soc.*, 2019, **141**, 5409–5414.

19  J. Lin, S. Zhang, W. Guan, G. Yang and Y. Ma, *J. Am. Chem. Soc.*, 2018, **140**, 9545–9550.

20  G. Liu, X. Feng, L. Wang, S. A. T. Redfern, X. Yong, G. Gao and H. Liu, *Phys. Chem. Chem. Phys.*, 2019, **21**, 17621–17627.

21  D. Kurzydłowski and W. Grochala, *Chem. Commun.*, 2008, **1**, 1073–1075.

22  D. Kurzydłowski and W. Grochala, *Z. Anorg. Allg. Chem.*, 2008, **634**, 1082–1086.

23  T. Söhnel, H. Hermann and P. Schwerdtfeger, *J. Phys. Chem. B*, 2005, **109**, 526–531.

24  B. Zemva, K. Lutar, A. Jesih, W. J. Casteel, A. P. Wilkinson, D. E. Cox, R. B. Von Dreele, H. Borrmann and N. Bartlett, *J. Am. Chem. Soc.*, 1991, **113**, 4192–4198.

25  R. D. Burbank and F. N. Bensey, *J. Chem. Phys.*, 1957, **27**, 982–983.

26  P. Schwerdtfeger, P. D. W. Boyd, S. Brienne and A. K. Burrell, *Inorg. Chem.*, 1992, **31**, 3411–3422.

27  J. Haines, J. M. Léger, F. Gorelli, D. D. Klug, J. S. Tse and Z. Q. Li, *Phys. Rev. B*, 2001, **64**, 134110.

28  D. Kurzydłowski, A. Oleksiak, S. B. Pillai and P. K. Jha, *Inorg. Chem.*, 2020, **59**, 2584–2593.

29  J. Linnera, S. I. Ivlev, F. Kraus and A. J. Karttunen, *Z. Anorg. Allg. Chem.*, 2019, **645**, 284–291.

30  A. Hermann, M. Derzsi, W. Grochala and R. Hoffmann, *Inorg. Chem.*, 2016, **55**, 1278–1286.

31  M. Tang, Y. Zhang, S. Li, X. Wu, Y. Jia and G. Yang, *ChemPhysChem*, 2018, **19**, 2989–2994.

32  J. Zhang, X. Feng, G. Liu, S. A. T. Redfern and H. Liu, *J. Phys. Condens. Matter*, 2020, **32**, 015402.

33  M. Wojdyr, *J. Appl. Crystallogr.*, 2010, **43**, 1126–1128.

34  A. Dewaele, M. Torrent, P. Loubeyre and M. Mezouar, *Phys. Rev. B*, 2008, **78**, 104102.

35  Y. Akahama and H. Kawamura, *J. Appl. Phys.*, 2006, **100**, 043516.

36  J. Sun, A. Ruzsinszky and J. P. Perdew, *Phys. Rev. Lett.*, 2015, **115**, 036402.

37  P. E. Blöchl, *Phys. Rev. B*, 1994, **50**, 17953–17979.

38  G. Kresse and J. Furthmüller, *Phys. Rev. B*, 1996, **54**, 11169–11186.

39  G. Kresse and D. Joubert, *Phys. Rev. B*, 1999, **59**, 1758–1775.

40  P. Avery, C. Toher, S. Curtarolo and E. Zurek, *Comput. Phys. Commun.*, 2019, **237**, 274–275.

41  J. P. Perdew, K. Burke and M. Ernzerhof, *Phys. Rev. Lett.*, 1996, **77**, 3865–3868.

42  K. Refson, P. R. Tulip and S. J. Clark, *Phys. Rev. B*, 2006, **73**, 155114.

43  S. J. Clark, M. D. Segall, C. J. Pickard, P. J. Hasnip, M. I. J. Probert, K. Refson and M. C. Payne, *Zeitschrift für Krist. - Cryst. Mater.*, 2005, **220**, 567–570.

44  E. E. Zvereva, A. R. Shagidullin and S. A. Katsyuba, *J. Phys. Chem. A*, 2011, **115**, 63–69.

45  A. V. Krukau, O. A. Vydrov, A. F. Izmaylov and G. E. Scuseria, *J. Chem. Phys.*, 2006, **125**, 224106.

46  K. Momma and F. Izumi, *J. Appl. Crystallogr.*, 2011, **44**, 1272–1276.

47  H. T. Stokes and D. M. Hatch, *J. Appl. Crystallogr.*, 2005, **38**, 237–238.

48  T. Björkman, *Comput. Phys. Commun.*, 2011, **182**, 1183–1186.

49  E. Kroumova, M. I. Aroyo, J. M. Perez-Mato, A. Kirov, C. Capillas, S. Ivantchev and H. Wondratschek, *Phase Transitions*, 2003, **76**, 155–170.

50  M. A. Ellwanger, S. Steinhauer, P. Golz, H. Beckers, A. Wiesner, B. Braun-Cula, T. Braun and S. Riedel, *Chem. - A Eur. J.*, 2017, **23**, 13501–13509.